# Intrapulse Impact Processes in Dense-Gas Femtosecond Laser Filamentation

Dmitri A. Romanov,[1, 4] Xiaohui Gao,[3] Alexander L. Gaeta,[3] and Robert J. Levis[2, 4]

[1]Department of Physics, Temple University, Philadelphia, PA 19122, USA
[2]Department of Chemistry, Temple University, Philadelphia, PA 19122, USA
[3]Department of Applied Physics and Applied Mathematics,
Columbia University, New York, NY 10027 USA
[4]Center for Advanced Photonics Research, College of Science and Technology,
Temple University, Philadelphia, PA 19122, USA

The processes of energy gain and redistribution in a dense gas subject to an intense ultrashort laser pulse are investigated theoretically for the case of high-pressure argon. The electrons released via strong-field ionization and driven by oscillating laser field collide with neutral neighbor atoms, thus effecting the energy gain in the emerging electron gas via a short-range inverse Bremsstrahlung interaction. These collisions also cause excitation and impact ionization of the atoms thus reducing the electron-gas energy. A kinetic model of these competing processes is developed which predicts the prevalence of excited atoms over ionized atoms by the end of the laser pulse. The creation of a significant number of excited atoms during the pulse in high-pressure gases is consistent with the delayed ionization dynamics in the pulse wake, recently discovered by Gao et al[1]. This energy redistribution mechanism offers an approach to manage effectively the excitation vs. ionization patterns in dense gases interacting with intense laser pulses and thus opens new avenues for diagnostics and control in these settings.

Strong-field ionization processes have been a hallmark of the physics of intense ultrashort laser pulses interacting with atoms and molecules. Considerable attention has been paid both to the details of the ionization event[2] and to the motion of the released electrons driven by the oscillating laser field.[3] One spectacular manifestation of these processes is high harmonic generation (HHG), which results from recombination of a field-accelerated electron with its parent ion, leading to emission of coherent soft X-rays and trains of attosecond pulses.[4,5] Recent research is exploring the more subtle effects of frustrated tunnel ionization[6] and so-called Kramers-Henneberger quasi-bound states in strong laser fields [7], which may affect HHG [8] and other nonlinear effects, such as higher-order Kerr effect, acceleration of neutral atoms, [9] and transient amplification of intense light fields in gases.[10] As a result of these processes, by the end of the laser pulse the low-momentum freed electrons may be partially recaptured by the ionic Coulomb potential into highly excited Rydberg states.

Strong-field ionization was first investigated in vacuum or at low pressure, where the laser field interacts with individual atoms or molecules. When strong-field ionization occurs in an atmospheric-pressure gas, many electrons are created per cubic wavelength, and their effect on laser pulse propagation becomes substantial. In particular, the emerging optically underdense plasma effectively counters the Kerr-effect self-focusing of the laser beam, leading to stabilized filamentation of femtosecond pulses.[11,12] Moreover, transient ionization grating in two-beam coupling settings may serve both for extracting the details of the ionization process[13-15] and for controlling subsequent laser pulses via interaction with the evolving grating.

In the wake of a filamenting laser pulse, the system of energetic free electrons, ions, and remaining neutral atoms (or molecules) evolves toward equilibrium, and this wake channel evolution gives rise to a number of transient effects which have both fundamental and practical importance. Post-pulse electron dynamics in the channel have been investigated using transverse interferometry,[16,17] four-wave mixing,[18,19] and measurement of Rabi sideband emission.[20,21] The results of these measurements agree well with the results of numerical simulations based on simple models of thermalized electron gas dynamics assisted by impact ionization cooling.[12,18,19] The free-electron density in a filament wake channel can be further controlled and manipulated by using avalanche ionization driven by an additional high-energy heater pulse of nanosecond duration.[22]

Further complications may be expected when strong-field ionization occurs in a dense (high-pressure) gas. Such high-pressure (and multiply ionized) noble gases were used to achieve efficient ultrahigh-harmonic generation, in which case the combination of linear response of electron gas and nonlinear response of ions was shown to engineer phase matching with the driving UV laser pulse.[5,23,24] Most recently, time-resolved transverse interferometry and shadowgraphy was used to trace evolution of the electron density in the filament wake in a high-pressure argon gas (60 bar) on picosecond time scale,[1] where an approximately 3-fold increase in the electron density was observed, saturating at about 30 picoseconds after the filamenting laser pulse.

Analysis of these observations[1] strongly suggests that the interaction of an intense femtosecond laser pulse with the dense gas should result in a density of excited argon atoms that is greater than that of generated ions. This high density of excited atoms cannot be accounted for by the Rydberg-state excitations resulting from the frustrated ionization, as mentioned above. First, at the typical value of laser intensity used in Ref. [1], $I = 7 \cdot 10^{13}$ W/cm², the calculated concentration of Rydberg atoms resulting from the frustrated ionization processes is less than 1%.[25] Second, the high-pressure gas is not conducive for the survival of high Rydberg states, because they immediately become subject to extensive autoionization. Therefore, the robust excited states in question should lie deep in the excited state manifold. The high density of moderately excited atoms also does not agree with the scenario of avalanche ionization, which is typical for longer laser pulses,[22] because the avalanche process produces a growing number of energetic electrons capable of impact-ionizing even the ground-state atoms, let alone the excited ones. Thus, the case of ultrashort laser pulse interacting with a high-pressure gas presents a distinct regime, which requires consideration of the physical mechanisms involved and of the system evolution towards the state obtained at the end of the ionizing laser pulse.

In this Communication, we investigate theoretically the evolution of a dense gas during interaction with an intense, ultrashort laser pulse. Similar to recent experimental work,[1] we assume an argon gas at a pressure of 60 atm. This gas interacts with a laser pulse at 800 nm, a duration of 70 fs (FWHM), and a focal-volume intensity in the range 0.5-1.0 × 10¹⁴ W/cm². We develop a kinetic model for the concerted evolution of the field-driven free electrons, the ions emerging from impact ionization, and the neutral atoms undergoing collisional excitation. As will be seen, the most important process in this evolution is the inverse Bremsstrahlung that occurs during short-range collisions with neutral atoms, which differs from the usual inverse Bremsstrahlung in extensively-ionized plasma that is due to long-range Coulomb interaction of electrons with ions.

Over this range of laser intensities, the laser-gas interaction leads to a moderate ionization rate, so when the released electrons are being driven by the oscillating laser field, they are most likely to collide with neutral atoms. These collisions cause the electron gas to gain energy via the short-range inverse Bremsstrahlung. As the electrons acquire energy in excess of their moderate ponderomotive energy value, the collisions can cause excitation and ionization of the atoms thus working to reduce the overall electron-gas energy. As will be shown, the evolution of electron gas driven by these competing processes leads to prevalence of the excited atoms over ionized atoms by the end of the laser pulse. The considerable build-up of excited atoms during interaction with the laser pulse in high-pressure gases explains the recently-discovered delayed ionization dynamics in the pulse wake.

In the cases of low- or atmospheric-pressure gases, the distance between the gas atoms/molecules is large enough ($\geq 30$ Å) to allow for the quivering motion of the released electron near its parent ion to be considered individually, separately from other atoms. In contrast, in the case of a high-pressure gas, the distance between the gas atoms becomes smaller than the ponderomotive radius ($\sim 8$ Å at the laser intensity of $0.5 \times 10^{14}$ W/cm$^2$), and thus collisions of the quivering electron with the neutral neighbors become significant. These collisions lead to fast dephasing of the initial electron wavepacket and make it possible to approximately consider the subsequent evolution of laser-driven electron gas in the kinetic equation framework.

The kinetics of the emerging electron gas during the driving laser pulse is comprised of three major processes: (i) energy gain via inverse Bremsstrahlung; (ii) impact excitation of neutral atoms, which is associated with energy loss by the free electrons; and (iii) impact ionization, which is associated with both the energy loss by free electrons and the generation of new free electrons. For simplicity, we consider a homogeneous electron gas and trace its evolution in terms of the density of occupied energy states $n(\varepsilon,t)$ so that the total electron number density is $n_{tot}(t) = \int_0^\infty d\varepsilon\, n(\varepsilon,t)$.

In constructing the inverse-Bremsstrahlung kinetics, we assume that the laser carrier frequency $\omega$ is greater than the elastic scattering rate of the ionized electrons with neutral atoms, which in turn is greater than the inverse of the pulse duration, $T$. The linearly polarized laser field exerts a force on a free electron, $\mathbf{F}(t) = \hat{\boldsymbol{\epsilon}} e E_0 s(t) \cos(\omega t)$, where $\hat{\boldsymbol{\epsilon}}$ is the laser polarization vector, $E_0$ is the laser field amplitude, and $s(t)$ is the dimensionless pulse envelope function. The electron distribution function $f(\mathbf{p},t)$ does not depend on spatial coordinates and satisfies the kinetic equation,

$$\frac{\partial f}{\partial t} + eE_0 s(t)\cos(\omega t)\hat{\boldsymbol{\epsilon}}\cdot\nabla_{\mathbf{p}} f = St\{f\}, \qquad (1)$$

where $St\{f\}$ is the collision integral, and $\nabla_{\mathbf{p}}$ is the gradient in the $\mathbf{p}$-space. In this equation, the laser field works to make $f(\mathbf{p},t)$ an anisotropic function of the momentum $\mathbf{p}$, while the collisions work to restore isotropy. Accordingly, it is convenient to separate explicitly the induced anisotropy and to represent approximately the distribution function in terms of two components: $f(\mathbf{p},t) = f_0(\varepsilon,t) + (\hat{\boldsymbol{\epsilon}}\cdot\mathbf{v}(\mathbf{p})) f_1(\varepsilon,t)$, where $\mathbf{v}(\mathbf{p}) = \mathbf{p}/m$ is the velocity vector. In this expression, the function $f_0(\varepsilon,t)$ represents the isotropic part of the distribution function, which depends on the electron energy, $\varepsilon$, and time $t$. The function $f_1(\varepsilon,t)$ represents the anisotropic part, as the angle averaging of the term $(\hat{\boldsymbol{\epsilon}}\cdot\mathbf{v}(\mathbf{p})) f_1(\varepsilon,t)$ identically produces zero. For the elastic short-range scattering from neutral atoms, the collision integral is expressed as,

$$St_p\{f\} = n_0 v^2(\varepsilon) f_1(\varepsilon) \times \\ \int d\Omega' \sigma(\varepsilon,\theta'-\theta)(\cos(\theta')-\cos(\theta)). \qquad (2)$$

In this expression, $n_0$ is the density of the neutral scatterers, $\sigma(\varepsilon,\theta)$ is the differential cross-section, $\theta'$ and $\theta$ are the angles between the vector $\mathbf{e}$ and the vectors $\mathbf{p}'$ and $\mathbf{p}$, respectively, the solid angle $\Omega'$ corresponds to the direction of vector $\mathbf{p}'$, and $v(\varepsilon) = p/m = \sqrt{2\varepsilon/m}$ is the electron velocity magnitude. Substitution of this expression for $St_p\{f\}$ in Eq. (1) and angle averaging of the resulting equation produces a relation between $f_0(\varepsilon,t)$ and $f_1(\varepsilon,t)$. A complementary relation between these functions may be obtained by angle averaging of the same resulting equation upon multiplying both sides of it by $\cos(\theta)$. After some algebra, the system of coupled partial differential equations for $f_0(\varepsilon,t)$ and $f_1(\varepsilon,t)$ is obtained in the form,

$$\begin{cases} \dfrac{\partial f_0}{\partial t} = -\dfrac{eE_0}{m}s(t)\cos(\omega t)\left(1+\dfrac{2}{3}\varepsilon\dfrac{\partial}{\partial \varepsilon}\right)f_1 \\ \dfrac{\partial f_1}{\partial t} = -\omega\beta(\varepsilon)f_1(\varepsilon)-eE_0 s(t)\cos(\omega t)\dfrac{\partial f_0}{\partial \varepsilon} \end{cases}, \quad (3)$$

where

$$\beta(\varepsilon) = \dfrac{n_0 v(\varepsilon)\sigma_{tr}(\varepsilon)}{\omega} = n_0\sigma_{tr}(\varepsilon)\sqrt{\dfrac{2\varepsilon}{m\omega^2}}; \quad (4)$$

$$\sigma_{tr}(\varepsilon) = \int d\Omega'' \sigma(\varepsilon,\theta'')(1-\cos(\theta'')),$$

$\sigma_{tr}(\varepsilon)$ being the transport cross-section. Solving formally the second equation in (3) for $f_1(\varepsilon,t)$ in terms of $f_0(\varepsilon,t)$, substituting this $f_1(\varepsilon,t)$ in the first equation in (3), and making use of the mentioned ranking of the characteristic frequencies, $\omega \gg \omega\beta \gg 1/T$, a closed equation for the laser-cycle-averaged function $\bar{f}_0(\varepsilon,t)$ is obtained as, $\partial \bar{f}_0/\partial t = 2\omega U_p s^2(t) \times (1+(2/3)\varepsilon\partial/\partial\varepsilon)(\beta(\varepsilon)\partial\bar{f}_0/\partial\varepsilon)$, where $U_p = (eE_0)^2/(4m\omega^2)$ is the ponderomotive potential. The function $\bar{f}_0(\varepsilon,t)$ relates to $n(\varepsilon,t)$ through the expression, $n(\varepsilon,t) = 2\pi(2m)^{3/2} n_{tot} \times \sqrt{\varepsilon}\bar{f}_0(\varepsilon,t)$ Finally, in the energy range of interest the transport cross-section is approximately constant, $\sigma_{tr}(\varepsilon) \approx \sigma_0 = 10^{-15}$ cm.[2,26] Given these approximations, the equation for the evolution of $n(\varepsilon,t)$ due to the inverse Bremsstrahlung process becomes,

$$\dfrac{\partial n}{\partial t} = \dfrac{4}{3}\sqrt{\dfrac{2}{m}}n_0\sigma_0 U_p s^2(t)\dfrac{\partial}{\partial \varepsilon}\left[\sqrt{\varepsilon}\left(\varepsilon\dfrac{\partial n}{\partial \varepsilon}-\dfrac{1}{2}n\right)\right]. \quad (5)$$

The process described by this Fokker-Plank-type equation can be seen as an effective diffusion of the electron density along the energy coordinate, with a time-dependent effective diffusion coefficient.

In the process of collisional excitation, a free electron promotes a neutral atom to an excited state with the excitation energy, $\varepsilon_{ex}$, losing an equivalent amount of kinetic energy. The twofold effect of this process on $n(\varepsilon,t)$ is as follows: the number of electrons at a given $\varepsilon$ is decreasing as some electrons move to the energy position of $\varepsilon-\varepsilon_{ex}$ due to the collisional excitation of neutral atoms; at the same time, new electrons arrive at $\varepsilon$ as they are being in turn shifted down from $\varepsilon+\varepsilon_{ex}$. Accordingly, the contribution of collisional excitation to the evolution of $n(\varepsilon,t)$ is described as,

$$\left.\dfrac{\partial n}{\partial t}\right|_{ex} = -\nu_{ex}(\varepsilon)n(\varepsilon,t)\Theta(\varepsilon-\varepsilon_{ex}) + \nu_{ex}(\varepsilon+\varepsilon_{ex})n(\varepsilon+\varepsilon_{ex},t), \quad (6)$$

where the two terms on the right-hand side correspond to the loss and gain in the electron population at the energy $\varepsilon$ just described. Of course, to be able to lose $\varepsilon_{ex}$, the electron should have the kinetic energy $\varepsilon > \varepsilon_{ex}$ in the first place, and this condition is expressed via the Heaviside step-function, $\Theta(\varepsilon-\varepsilon_{ex})$. The collisional excitation rate, $\nu_{ex}(\varepsilon)$, is determined by the concentration of the neutral atoms, the electron velocity, and the energy-dependent total excitation cross-section, $\sigma_{ex}(\varepsilon)$, in the standard way: $\nu_{ex}(\varepsilon) = n_0\sigma_{ex}(\varepsilon)\sqrt{2\varepsilon/m}$. Strictly speaking, there are several accessible excited atomic states, so that the right-hand side in Eq. (6) should have contributions from transitions related to all of these states with the respective excitation rates. However, we will restrict our model with one representative excited state, which has the excitation energy of $\varepsilon_{ex} = 11.8$ eV. Furthermore, for $\tilde{\sigma}_{ex}(\varepsilon)$ we use the semi-empirical formula of Ref. [27], which reads,

$$\sigma_{ex}(\varepsilon) = 1.40 \pi a_B^2 \left(\frac{Ry}{\varepsilon_{ex}}\right)^2 \times \left(\frac{\varepsilon_{ex}}{\varepsilon}\right)^{0.75} \left(1 - \frac{\varepsilon_{ex}}{\varepsilon}\right)^2 \Theta(\varepsilon - \varepsilon_{ex}). \quad (7)$$

As a function of energy, $\sigma_{ex}(\varepsilon)$ has a threshold at $\varepsilon = \varepsilon_{ex}$, and near this threshold it rises according to a power law.

Along with its contribution to the evolution of $n(\varepsilon,t)$, each act of collisional excitation leads to emergence of an additional excited atom. Thus, the growth of the excited-atom density, $n_{ex}$, is determined by the first term in the right-hand side of Eq. (6), according to

$$\frac{dn_{ex}}{dt} = \int_{\varepsilon_{ex}}^{\infty} d\varepsilon \, v_{ex}(\varepsilon) n(\varepsilon,t). \quad (8)$$

The effects of impact ionization processes on $n(\varepsilon,t)$ can be described analogously to the effects of collisional excitation, as expressed in Eq. (6). However, one important distinction is that the total number of free electrons is not conserved, since each event of impact ionization adds a free electron. Accordingly, the contribution of the impact-ionization processes to the evolution of $n(\varepsilon,t)$ is expressed as,

$$\left.\frac{\partial n}{\partial t}\right|_{ion} = -v_{ion}(\varepsilon) n(\varepsilon,t) \Theta(\varepsilon - \varepsilon_{ion}) + v_{ion}(\varepsilon + \varepsilon_{ion}) n(\varepsilon + \varepsilon_{ion}, t) + g(\varepsilon) \int_{\varepsilon_{ion}}^{\infty} d\varepsilon \, v_{ion}(\varepsilon) n(\varepsilon,t) \quad , (9)$$

where $v_{ion}(\varepsilon) = n_0 \sigma_{ion}(\varepsilon) \sqrt{2\varepsilon/m}$ is the impact ionization rate, $\varepsilon_{ion}$ is the ionization energy, and $\sigma_{ion}(\varepsilon)$ is the energy-dependent total ionization cross-section. On the right-hand side of Eq. (9), the first and the second terms are analogous to the respective terms in Eq. (6), whereas the last term accounts for the free electrons emerging from the impact ionization. The growing total number of these secondary electrons is given by the integral, which is analogous to that of Eq. (8), and their energy distribution is modeled by an auxiliary function, $g(\varepsilon)$. (The electrons emerge with low kinetic energy, but upon the first cycle of laser-field acceleration and elastic scattering, which is beyond our model framework, they acquire some initial energy distribution.) The specific form of this function is of no particular importance, we choose $g(\varepsilon) = C_{m\mu} \varepsilon^m (1 - \tanh(\mu(\varepsilon - U_p)))$, where $m$ and $\mu$ are adjustable parameters and $C_{m\mu}$ is the normalization constant. The total number of ionized atoms grows in step with the total number of free electrons, and in direct analogy to Eq. (8) it is expressed as,

$$\frac{dn_{ion}}{dt} = \int_{\varepsilon_{ion}}^{\infty} d\varepsilon \, v_{ion}(\varepsilon) n(\varepsilon,t). \quad (10)$$

For the total ionization cross-section, $\sigma_{ion}(\varepsilon)$, we use the semi-empirical Lotz formula.[28] The cross-section is given by the sum of subshell contributions, and for Ar and for the electron energies in question, it is sufficient to consider only the contribution of the eight electrons of the uppermost shell:

$$\sigma_{ion}(\varepsilon) = \frac{a}{\varepsilon_{ion}\varepsilon} \cdot \Theta(\varepsilon - \varepsilon_{ion}) \times \left(1 - b \cdot \exp\left(-c\frac{\varepsilon - \varepsilon_{ion}}{\varepsilon_{ion}}\right)\right) \ln\left(\frac{\varepsilon}{\varepsilon_{ion}}\right) \quad (11)$$

where the empirical constants are: $a = 32 \cdot 10^{-14}$ cm$^2$ eV$^2$, $b = 0.62$, $c = 0.40$, and the ionization energy is $E_{ion} = 15.76$ eV. The function $\sigma_{ion}(\varepsilon)$ has a threshold at $\varepsilon_{ion}$, and then it rises sharply throughout the range of energies pertinent to the $n(\varepsilon,t)$ evolution.

To put Eqs. (5)-(11) in dimensionless variables, the energy unit is chosen to be the ponderomotive potential at the laser intensity of $I = 7 \cdot 10^{13}$ W/cm$^2$ and the wavelength of 800 nm: $\varepsilon_0 = 4.183$ eV. Then, the dimensionless excitation and ionization energies are $\varepsilon_{ex} \approx 2.77$ and $\varepsilon_{ion} \approx 3.77$, respectively. The time unit is convenient to choose as $t_0 = 3\sqrt{m}/(4n_0\sigma_0\sqrt{2\varepsilon_0})$, for which the estimates give the value $t_0 = 0.344 \cdot 10^{-14}$ s. With the duration of the laser pulse is assumed to be about 70 fs, the dimensionless time variable, $\tau = t/t_0$, ranges from $\tau = 0$ to about $\tau = 24$.

Upon combining Eqs. (5), (6), and (9), and using the introduced dimensionless variables $\varepsilon$ and $\tau$, the equation describing the evolution of $n(\varepsilon,t)$ is obtained in the dimensionless form as,

$$\frac{\partial n}{\partial \tau} = u_p s^2(\tau) \frac{\partial}{\partial \varepsilon}\left(\frac{\partial}{\partial \varepsilon}(\varepsilon^{3/2} n) - 2\varepsilon^{1/2} n\right) + \\ \tilde{\sigma}_{ex}(\varepsilon + \varepsilon_{ex})\sqrt{\varepsilon + \varepsilon_{ex}} n(\varepsilon + \varepsilon_{ex}) - \\ \tilde{\sigma}_{ex}(\varepsilon)\sqrt{\varepsilon} n(\varepsilon)\Theta(\varepsilon - \varepsilon_{ex}) + \\ \tilde{\sigma}_{ion}(\varepsilon + \varepsilon_{ion})\sqrt{\varepsilon + \varepsilon_{ion}} n(\varepsilon + \varepsilon_{ion}) - \\ \tilde{\sigma}_{ion}(\varepsilon)\sqrt{\varepsilon} n(\varepsilon)\Theta(\varepsilon - \varepsilon_{ion}) + \\ g(\varepsilon)\int_{\varepsilon_{ion}}^{\infty} d\varepsilon \, \tilde{\sigma}_{ion}(\varepsilon)\sqrt{\varepsilon} n(\varepsilon) + g(\varepsilon) W(\tau). \quad (12)$$

where $\tilde{\sigma}_{ex}(\varepsilon) = (3/4)(\sigma_{ex}(\varepsilon)/\sigma_0)$, $\tilde{\sigma}_{ion}(\varepsilon) = (3/4)(\sigma_{ion}(\varepsilon)/\sigma_0)$, and $W(t)$ is the rate of strong-field ionization by the laser pulse. In Eq. (12), the dimensionless function $n(\varepsilon,\tau)$ is normalized to $n_0$, and the dimensionless parameter $u_p$ is the ponderomotive potential normalized to $\varepsilon_0$. The equation describes the scenario in which the electrons gain energy through the inverse Bremsstrahlung process and spend that energy on excitation and ionization of the surrounding neutral atoms. The outcome of these competing processes depends essentially on the relative values and functional dependencies of $\tilde{\sigma}_{ex}(\varepsilon)$ and $\tilde{\sigma}_{ion}(\varepsilon)$, which determine the collisional excitation rate $\nu_{ex}(\varepsilon)$ and the ionization rate $\nu_{ion}(\varepsilon)$, respectively.

It is instructive to compare the energy dependence of $\nu_{ex}(\varepsilon)$ and $\nu_{ion}(\varepsilon)$, see Fig. 1. The curve of $\nu_{ion}(\varepsilon)$ rises faster than the curve of $\nu_{ex}(\varepsilon)$, and at sufficiently large energy of the colliding electron the impact ionization becomes preferable to the collisional excitation. However, as the value of $\varepsilon_{ex}$ is smaller than the value of $\varepsilon_{ion}$, there is a considerable energy interval in which $\tilde{\sigma}_{ex}(\varepsilon) > \tilde{\sigma}_{ion}(\varepsilon)$. The existence of this interval of predominant excitation reflects decisively on the evolution of the electron gas during the laser pulse. When the electrons climb up the energy scale, being driven by the inverse Bremsstrahlung process, they first come across this interval of preferential excitation and do actively excite the neutral atoms. Since the acts of excitation remove energy from the free electrons, the kinetic energy distribution shifts toward lower energies. From those lower energies, the electrons again climb up, only to get across the same interval of preferential excitation. Due to this Sisyphean cycling, the electrons hardly get through to the energies where they can effectively participate in impact ionization. As a result, the emerging electron energy distribution peaks at the energies just below $\varepsilon_{ex}$; this pattern will be seen in the numerical simulation presented in Fig. 2. With this scenario, we can expect a large number of excited neutral atoms to result from the laser-gas interaction, as compared with the resulting number of ions.

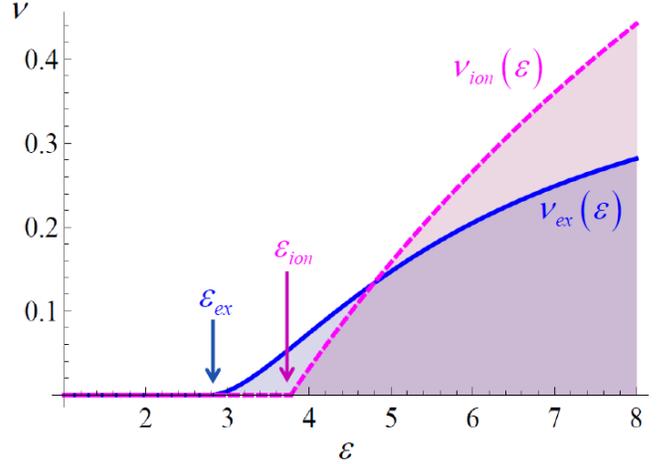

**Fig. 1**. (Color online) The rates of collisional excitation (blue-solid line) and the impact ionization (magenta-dashed line) as functions of the electron energy, effected by the respective total cross-sections presented by Eq. (7) and Eq. (11) in the text. $\nu$ is in units of $t_0^{-1}$ and $\varepsilon$ is in units of $\varepsilon_0$.

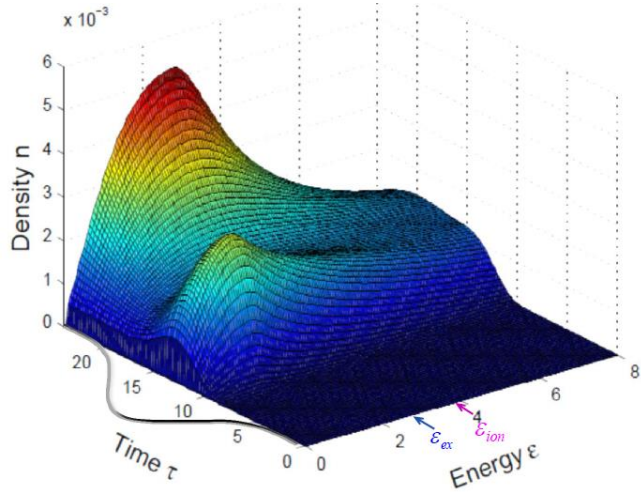

**Fig. 2**. (Color online) Evolution of the electron density distribution over energy during the laser pulse. The dimensionless energy and time variables are described in the paragraph after Eq. (11). The pulse envelope is modeled by cosine-square function centered at $\tau = 12$; the time-dependent laser intensity is shown symbolically as the gray curve. The initial rise of the electron density is due to strong-field ionization, the following build-up is due to impact processes energized by inverse Bremsstrahlung.

From mathematical standpoint, Eq. (12) is a complicated partial differential equation, which contains an integral term and finite-difference terms. As such, it is not amenable to standard numerical solvers and required a tailored approach. We have implemented an approach based the solution for delayed differential equations developed in Ref. [29] We use the model envelope function, $s(\tau) = \cos^2\left((\pi/T)(\tau - T/2)\right) \times \Theta(\tau)\Theta(T-\tau)$ with $T = 24$, model the strong-field ionization rate $W(t)$ using the ADK formula,[30] and solve Eq. (12) numerically for the evolution of $n(\varepsilon, \tau)$ during the laser pulse. The results are presented in Fig. 2.

The hump on the $n(\varepsilon, \tau)$ surface that is observed around $\tau = 12$ reflects the production of primary electrons by the laser pulse via strong-field ionization. As seen, it is dwarfed by the second hump, which is due to the impact ionization. As also seen, this second hump is effectively confined to the energies lower than the thresholds of excitation and ionization processes. The evolution of the total number densities of the excited atoms, $n_{ex}(\tau)$ and the ionized atoms $n_{ion}(\tau)$ during the laser pulse are obtained using Eq. (8) and Eq. (10), respectively, and are presented in Fig. 3. As seen, the ion density first rises sharply near the peak of the laser pulse (around $\tau = 12$), this occurs due to the strong-field ionization and corresponds to the mentioned first hump on the $n(\varepsilon, \tau)$ surface in Fig. 2. At later time, the ion density continues to grow steadily, now because of the production of secondary electrons via impact ionization. The density of the excited atoms starts to grow only after the peak of the laser pulse, which is quite natural, as the excitation is effected by energetic free electrons. At first, this density of excited atoms trails noticeably the number of ions, because near the peak of the laser pulse the primary electrons (and thus the primary ions) are produced extensively. However, at later times the excited-atoms curve rises much more sharply than that of the ions, and by the end of the laser pulse, the number of excited atoms is almost threefold larger than the number of ions, which is in good agreement with experimental observations.[1] Thus, our theoretical results confirm that in the case of a dense gas, a filamenting laser pulse can produce considerably greater density of excited atoms than the density of ions, which should play a decisive role in the after-pulse evolution of the filament wake channel.

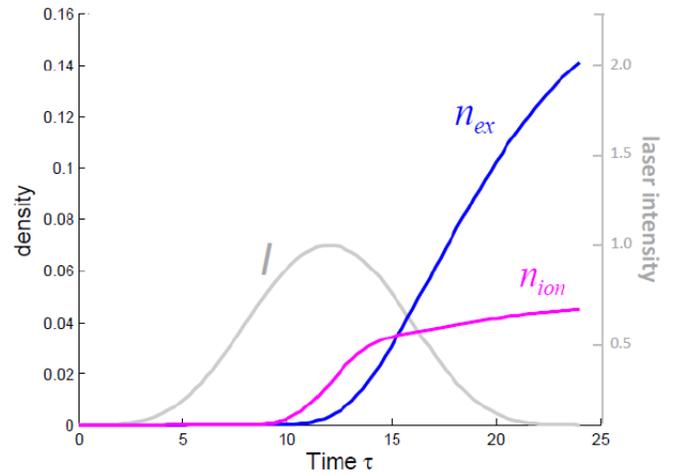

**Fig. 3**. (Color online) The total density of the ionized atoms (magenta curve) and the excited atoms (blue curve) vs. time during the laser pulse. The normalized laser intensity is shown for the reference as the gray curve. The dimensionless time variable is described in the paragraph after Eq. (11).

In conclusion, we have theoretically investigated the evolution of a dense argon gas being subjected to a strong-field, ultrashort laser pulse. We found that this evolution differs markedly from what happens in low-pressure gases, because the primary electrons, released by the strong-field ionization and driven by the laser electric field in a quivering motion around their parent ions, are very likely to collide with the neighboring neutral atoms. Due to these short-range collisions, an inverse-Bremsstrahlung process leads to effective energy gain by the electron gas. However, a considerable portion of this energy is channeled to inelastic collisional processes of impact ionization and collisional excitation of neutral atoms. Having lower energy threshold, the second process dominates over the first one, and leads to considerable build-up of excited atoms by the end of the laser pulse. The number of ions resulting from the impact ionization, the prevalence of the excited atoms, and the nonequilibrium electron distribution function set the stage for the longer-term system evolution after the laser pulse, which in turn determines the evolution the transient modifications of the linear and nonlinear optical properties of the wake channel. Also, the results obtained may be extended to the cases of lower-pressure gases interacting with mid-infrared lasers as the ponderomotive radius and thus the significance of collisional processes will scale as square of the laser wavelength.

The authors gratefully acknowledge support of the Office of Naval Research (N00014-15-1-2574); A.L.G. and X.G gratefully acknowledge support under a MURI grant from the Air Force Office of Scientific Research under Award No. FA9550-16-1-0121.

**References**

1. X. H. Gao, G. Patwardhan, S. Schrauth *et al.* Picosecond ionization dynamics in femtosecond filaments at high pressures, *Phys. Rev. A* **95**, Art. No. 013412 (5) (2017).
2. P. Agostini & L. F. DiMauro. Atomic and Molecular Ionization Dynamics in Strong Laser Fields: From Optical to X-rays, in *Advances in Atomic, Molecular, and Optical Physics, Vol 61* Vol. 61 (eds E. Arimondo, P. R. Berman, & C. C. Lin) 117 (Elsevier Academic Press Inc, 2012).
3. P. B. Corkum. Plasma Perspective on Strong-Field Multiphoton Ionization, *Phys. Rev. Lett.* **71**, 1994 (1993).
4. P. M. Paul, E. S. Toma, P. Breger *et al.* Observation of a train of attosecond pulses from high harmonic generation, *Science* **292**, 1689 (2001).
5. T. Popmintchev, M. C. Chen, D. Popmintchev *et al.* Bright Coherent Ultrahigh Harmonics in the keV X-ray Regime from Mid-Infrared Femtosecond Lasers, *Science* **336**, 1287 (2012).
6. T. Nubbemeyer, K. Gorling, A. Saenz, U. Eichmann & W. Sandner. Strong-Field Tunneling without Ionization, *Phys. Rev. Lett.* **101**, Art. No. 233001 (2008).
7. F. Morales, M. Richter, S. Patchkovskii & O. Smirnova. Imaging the Kramers-Henneberger atom, *Proc. Natl. Acad. Sci. U. S. A.* **108**, 16906 (2011).
8. T. Bredtmann, S. Chelkowski, A. D. Bandrauk & M. Ivanov. XUV lasing during strong-field-assisted transient absorption in molecules, *Phys. Rev. A* **93**, Art. No. 021402 (2016).
9. M. Richter, S. Patchkovskii, F. Morales, O. Smirnova & M. Ivanov. The role of the Kramers-

Henneberger atom in the higher-order Kerr effect, *New Journal of Physics* **15**, Art. No. 083012 (2013).
10. M. Matthews, F. Morales, A. Patas *et al.* Amplification of intense light fields by nearly free electrons, *Nature Physics* (2018).
11. A. Braun, G. Korn, X. Liu *et al.* Self-Channeling of High-Peak-Power Femtosecond Laser Pulses in Air, *Opt. Lett.* **20**, 73 (1995).
12. D. A. Romanov & R. J. Levis. Evolution of laser microfilaments in the wake of a femtosecond driving pulse, *Phys. Rev. A* **87**, Art. No. 063410 (8) (2013).
13. J. K. Wahlstrand & H. M. Milchberg. Effect of a plasma grating on pump-probe experiments near the ionization threshold in gases, *Opt. Lett.* **36**, 3822 (2011).
14. J. K. Wahlstrand, Y. H. Chen, Y. H. Cheng, S. R. Varma & H. M. Milchberg. Measurements of the High Field Optical Nonlinearity and Electron Density in Gases: Application to Filamentation Experiments, *IEEE J. Quantum Electron.* **48**, 760 (2012).
15. J. H. Odhner, D. A. Romanov, E. T. McCole *et al.* Ionization-Grating-Induced Nonlinear Phase Accumulation in Spectrally Resolved Transient Birefringence Measurements at 400 nm, *Phys. Rev. Lett.* **109**, Art. No. 065003 (5) (2012).
16. V. V. Bukin, S. V. Garnov, A. A. Malyutin & V. V. Strelkov. Interferometric diagnostics of femtosecond laser microplasma in gases, *Phys. Wave Phenom.* **20**, 91 (2012).
17. Y. H. Chen, S. Varma, T. M. Antonsen & H. M. Milchberg. Direct Measurement of the Electron Density of Extended Femtosecond Laser Pulse-Induced Filaments, *Phys. Rev. Lett.* **105**, Art. No. 215005 (4) (2010).
18. A. Filin, R. Compton, D. A. Romanov & R. J. Levis. Impact-Ionization Cooling in Laser-Induced Plasma Filaments, *Phys. Rev. Lett.* **102**, Art. No. 155004 (4) (2009).
19. D. A. Romanov, R. Compton, A. Filin & R. J. Levis. Dynamics of strong-field laser-induced microplasma formation in noble gases, *Phys. Rev. A* **81**, Art. No. 033403 (8) (2010).
20. R. Compton, A. Filin, D. A. Romanov & R. J. Levis. Observation of Broadband Time-Dependent Rabi Shifting in Microplasmas, *Phys. Rev. Lett.* **103**, Art. No. 205001 (4) (2009).
21. R. Compton, A. Filin, D. A. Romanov & R. J. Levis. Dynamic Rabi sidebands in laser-generated microplasmas: Tunability and control, *Phys. Rev. A* **83**, Art. No. 053423 (8) (2011).
22. P. Polynkin, B. Pasenhow, N. Driscoll *et al.* Seeded optically driven avalanche ionization in molecular and noble gases, *Phys. Rev. A* **86**, Art. No. 043410 (8) (2012).
23. T. Popmintchev, M. C. Chen, A. Bahabad *et al.* Phase matching of high harmonic generation in the soft and hard X-ray regions of the spectrum, *Proc. Natl. Acad. Sci. U. S. A.* **106**, 10516 (2009).
24. D. Popmintchev, C. Hernandez-Garcia, F. Dollar *et al.* Ultraviolet surprise: Efficient soft x-ray high-harmonic generation in multiply ionized plasmas, *Science* **350**, 1225 (2015).
25. H. Zimmermann, S. Patchkovskii, M. Ivanov & U. Eichmann. Unified Time and Frequency Picture of Ultrafast Atomic Excitation in Strong Laser Fields, *Phys. Rev. Lett.* **118**, Art. No. 013003 (2017).
26. E. Gargioni & B. Grosswendt. Electron scattering from argon: Data evaluation and consistency, *Rev. Mod. Phys.* **80**, 451 (2008).
27. L. R. Peterson & J. E. Allen. Electron-Impact Cross-Sections for Argon, *J. Chem. Phys.* **56**, 6068 (1972).
28. W. Lotz. An Empirical Formula for Electron-Impact Ionization Cross-Section, *Zeitschrift Fur Physik* **206**, 205 (1967).
29. L. F. Shampine & S. Thompson. Solving DDEs in MATLAB, *Appl. Numer. Math.* **37**, 441 (2001).
30. M. V. Ammosov, N. B. Delone & V. P. Krainov. Tunnel Ionization of Complex Atoms and Atomic Ions in a Varying Electromagnetic Field, *Zhurnal Eksperimentalnoi Teor. Fiz.* **91**, 2008 (1986).